\documentclass[aps,epsfig,prb,preprint]{revtex4}

\usepackage{amsmath}
\usepackage{graphicx}

\def\nn{\nonumber}

\begin{document}

\title{Decay and Frequency Shift of Inter and Intravalley Phonons
in Graphene $-$Dirac Cone Migration$-$}

\author{Ken-ichi Sasaki}
\email{sasaki.kenichi@lab.ntt.co.jp}
\affiliation{NTT Basic Research Laboratories, 
Nippon Telegraph and Telephone Corporation,
3-1 Morinosato Wakamiya, Atsugi, Kanagawa 243-0198, Japan}

\author{Keiko Kato}
\affiliation{NTT Basic Research Laboratories, 
Nippon Telegraph and Telephone Corporation,
3-1 Morinosato Wakamiya, Atsugi, Kanagawa 243-0198, Japan}

\author{Yasuhiro Tokura}
\affiliation{NTT Basic Research Laboratories, 
Nippon Telegraph and Telephone Corporation,
3-1 Morinosato Wakamiya, Atsugi, Kanagawa 243-0198, Japan}

\author{Satoru Suzuki}
\affiliation{NTT Basic Research Laboratories, 
Nippon Telegraph and Telephone Corporation,
3-1 Morinosato Wakamiya, Atsugi, Kanagawa 243-0198, Japan}

\author{Tetsuomi Sogawa}
\affiliation{NTT Basic Research Laboratories, 
Nippon Telegraph and Telephone Corporation,
3-1 Morinosato Wakamiya, Atsugi, Kanagawa 243-0198, Japan}

\date{\today}

\begin{abstract}
 By considering analytical expressions for 
 the self-energies of intervalley and intravalley phonons in graphene,
 we describe the behavior of D, 2D, and D$'$ Raman bands 
 with changes in doping ($\mu$) and light excitation energy ($E_L$).
 Comparing the self-energy with 
 the observed $\mu$ dependence of the 2D bandwidth,
 we estimate the wavevector $q$ of the constituent intervalley phonon 
 at $\hbar vq\simeq E_L/1.6$ ($v$ is electron's Fermi velocity) and 
 conclude that the self-energy makes a major contribution (60\%) 
 to the dispersive behavior of the D and 2D bands.
 The estimation of $q$ is based on a concept of shifted Dirac cones 
 in which the resonance decay of a phonon satisfying $q >  \omega/v$
 ($\omega$ is the phonon frequency) 
 into an electron-hole pair is suppressed when $\mu < (\hbar vq-\hbar\omega)/2$.
 We highlight the fact that
 the decay of an intervalley (and intravalley longitudinal optical) 
 phonon with $q=\omega/v$ 
 is strongly suppressed by electron-phonon
 coupling at an arbitrary $\mu$.
 This feature is in contrast to the divergent behavior of an
 intravalley transverse optical phonon, 
 which bears a close similarity to the
 polarization function relevant to plasmons.
\end{abstract}

\maketitle

The Raman spectrum of graphene has two prominent peaks 
called the G and 2D (or G$'$) bands that are very informative
characterization tools.
The 2D band at $\sim$ 2600 cm$^{-1}$ has been used to distinguish a
single layer from graphene layers.~\cite{ferrari06}
The G band at $\sim$ 1580 cm$^{-1}$ can be used to determine 
whether or not the position of the
Fermi energy $\mu$ is close to the Dirac point,
since the width broadens when $\mu \simeq 0$.~\cite{yan07,pisana07,das08nature,lazzeri06prl,ando06-ka}
By contrast, the 2D bandwidth sharpens when $\mu \simeq 0$.~\cite{das08nature,chen11}
What is the origin of the difference between the $\mu$ dependencies of
the G and 2D bands?

\begin{figure}[htbp]
 \begin{center}
  \includegraphics[scale=0.65]{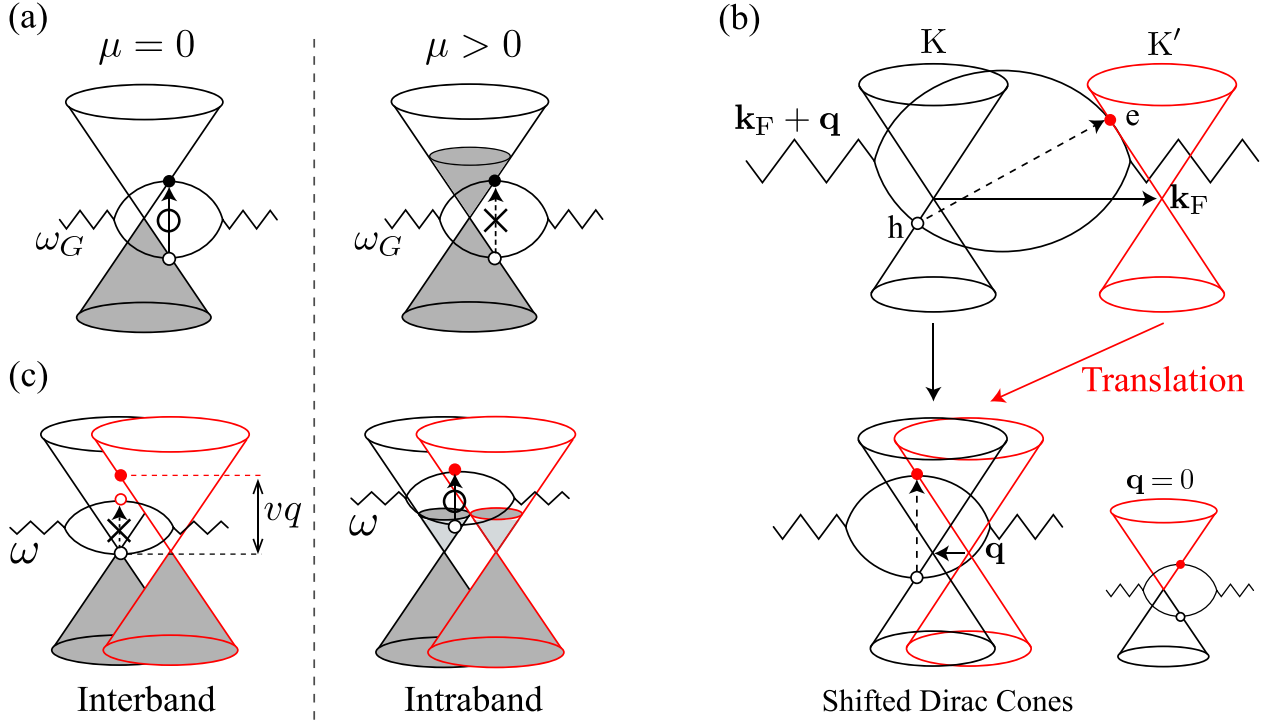}
 \end{center}
 \caption{
 (a) The $\mu$ dependence of G band broadening (frequency $\omega_G$).
 (b) An electron-hole pair between two valleys. 
 The Dirac cone at the K (K$'$) point is indicated in black (red).
 With the migration of the Dirac cone,
 the electron-hole pair creation process is viewed as a direct
 transition.
 (c)
 The $\mu$ dependence of the broadening of an intervalley phonon is
 different from that of the G band. 
 Note that 
 spectral broadening of an intravalley phonon
 can be discussed with the replacement ${\rm K}' \to {\rm K}$
 (or ${\rm K} \to {\rm K}'$).
 }
 \label{fig:migration}
\end{figure}

As illustrated in Fig.~\ref{fig:migration}(a),
the presence (absence) of a resonant process by which 
the phonon decays into a real electron-hole pair, 
enhances (suppresses) the spectral broadening.
Because the G band consists of $\Gamma$ point phonons, 
a direct transition is a unique decay channel 
that conserves momentum.
Thus, the $\mu$ dependence of the G bandwidth is readily understood
in terms of the Pauli exclusion
principle.~\cite{yan07,pisana07,das08nature,lazzeri06prl,ando06-ka}  
Meanwhile, the 2D band involves two near K point (intervalley)
phonons,~\cite{thomsen00,saito02prl} and the spectral broadening is induced by an
indirect transition that crosses two valleys, as shown in Fig.~\ref{fig:migration}(b).
The presence or absence of a resonance decay channel for a phonon with
a nonzero wavevector is the key to answering the question posed above.
In this paper, 
we provide the answer in a unified manner by translating the Dirac cone.

Figure~\ref{fig:migration}(b) shows that 
an intervalley phonon (zigzag line) 
can change into an electron-hole pair (loop) as a result of an
electron-phonon interaction.
The wavevector of an intervalley phonon
is written as ${\bf k}_{\rm F}+{\bf q}$,
where ${\bf k}_{\rm F}$ is a wavevector 
pointing from the K point to the K$'$ point and 
$q$ $(=|{\bf q}|)$ is much smaller than $|{\bf k}_{\rm F}|$.
Suppose that a hole is located at ${\bf k}$
measured from the K point, 
then the wavevector of the electron is given by 
${\bf k}+ ({\bf k}_{\rm F}+{\bf q})$
because of momentum conservation.
As a result, the wavevector of the electron 
measured from the K$'$ point is ${\bf k}+{\bf q}$, and 
the energies of the hole and electron are given by
$-\hbar vk$ ($=-\hbar v|{\bf k}|$) and $\hbar v|{\bf k+q}|$,
respectively, where $v$ ($\sim 10^6$ m/s) 
denotes the Fermi velocity.
Hereafter, we use units in which $\hbar =1$.

An indirect transition
between two valleys can be regarded
as a ``direct'' transition 
by translating the Dirac cone at the K$'$ point 
to $- ({\bf k}_{\rm F}+{\bf q})$ 
as shown in Fig.~\ref{fig:migration}(b).
With the shifted Dirac cones, 
it is easy to capture the essential feature of
the broadening of a ${\bf q}\ne 0$ phonon.
When $\mu=0$,
we see in Fig.~\ref{fig:migration}(c) that
there is an energy gap, $vq$, between the conduction and valence bands.
This energy gap precludes a phonon with frequency $\omega$ ($<vq$)
from decaying into a real electron-hole pair.
On the other hand, 
when sufficient doping is achieved
as shown in Fig.~\ref{fig:migration}(c), 
the phonon can decay into an intraband electron-hole pair.
This intraband decay channel 
results in spectral broadening.
When $q=0$, the two Dirac cones are merged into one [inset
in Fig.~\ref{fig:migration}(b)] and the energy gap vanishes.
Then, it is clear that the broadening of the ${\bf q}= 0$ phonon bears
similarities to that of the G
band.~\cite{yan07,pisana07,das08nature,lazzeri06prl,ando06-ka}  
The $\mu$ dependence of the broadening of an intervalley phonon
with $vq>\omega$ differs greatly from that of the G band, and
the concept of the shifted Dirac cones is useful for understanding the
$\mu$ and $q$ dependencies of the broadening in a unified manner.

More detailed information about the broadening 
can be obtained by calculating the self-energy.
The self-energy of the intervalley phonon with ${\bf q}$ and 
$\omega$ ($>0$) is defined by
\begin{align}
 \Pi_\mu(q,\omega) \equiv 
 g_{\rm ep} \frac{(2\pi)^2}{V}\sum_{s,s'} \sum_{\bf k} 
 \frac{f^{s}_{{\bf k},\mu}-f^{s'}_{{\bf k+q},\mu}}{\omega+svk-s'v|{\bf k+q}|+i\epsilon}
 \left( 1-ss'\frac{k+q\cos\varphi}{|{\bf k+q}|} \right).
 \label{eq:Pi}
\end{align}
In Eq.~(\ref{eq:Pi}), $s$ ($=\pm1$) and $s'$ ($=\pm1$) are band indices, 
$f^s_{{\bf k},\mu}=\lim_{\beta\to\infty}
(1+e^{\beta (s v|{\bf k}|-\mu)})^{-1}$ 
is the Fermi distribution function
defined at zero temperature 
and with a finite doping $\mu$,
and $\epsilon$ is a positive infinitesimal.
We can assume $\mu \ge 0$ without losing generality 
because of particle-hole symmetry.
The factor $g_{\rm ep}$ denotes the electron-phonon coupling strength,
$\varphi$ denotes the polar angle between ${\bf k}$ and ${\bf q}$,
and the term, $g_{\rm ep}\times(1-ss'\frac{k+q\cos\varphi}{|{\bf k+q}|})$, 
is the square of the electron-phonon matrix element 
for the intervalley phonon, which will be discussed later.
The broadening and modified frequency are given by
$-{\rm Im} \Pi_\mu(q,\omega)$ and 
$\omega + {\rm Re} \Pi_\mu(q,\omega)$, respectively.

\begin{figure}[htbp]
 \begin{center}
  \includegraphics[scale=0.7]{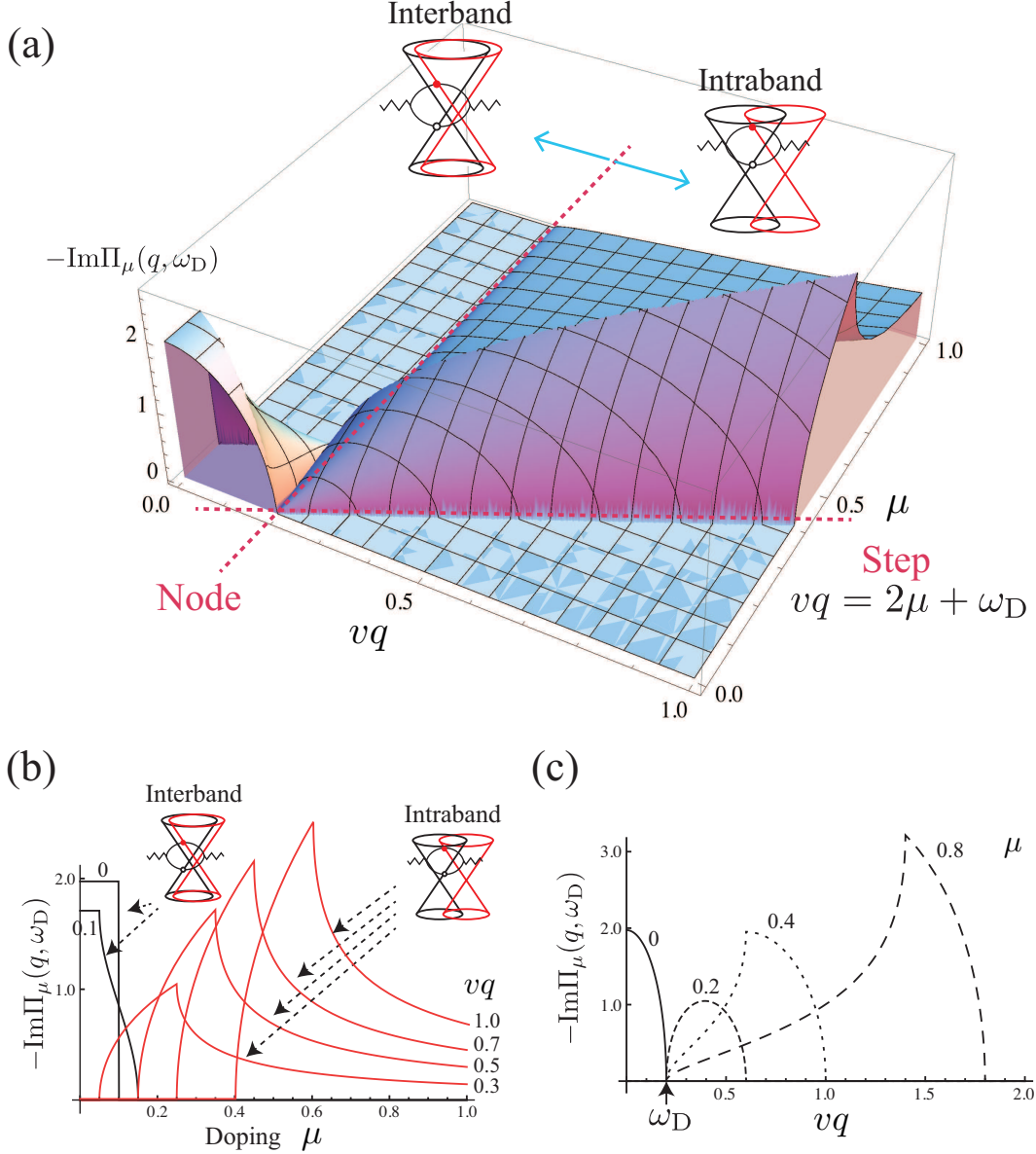}
 \end{center}
 \caption{
 (a) 3d plot of $-{\rm Im}\Pi_\mu(q,\omega_D)$.
 The variables $vq$ and $\mu$ are given in units of eV. 
 The cross section of (a) for different $vq$ values (b), 
 and for different $\mu$ values (c). 
 In (c), $vq$ is proportional to the light excitation energy $E_L$.
 }
 \label{fig:imPI}
\end{figure}

In the continuum limit of ${\bf k}$, 
the broadening normalized by $g_{\rm ep}$ leads to~\footnote{See Supplemental Material at [URL] for derivation.}
\begin{align}
 -{\rm Im} \Pi_\mu(q,\omega) =
 & \pi
 \sqrt{\omega^2-v^2q^2} \theta_{\omega-vq} \left[
 \theta_{\frac{\omega-vq}{2}-\mu} \pi
 + \theta_{\mu-\frac{\omega-vq}{2}} \theta_{\frac{\omega+vq}{2}-\mu}
 \left\{ 
 \frac{\pi}{2}-\sin^{-1}\left(\frac{2\mu-\omega}{vq}\right) \right\}
 \right]  \nn \\
 &+ \pi
 \sqrt{v^2q^2-\omega^2} \theta_{vq-\omega}
 \left[  \theta_{\mu-\frac{vq-\omega}{2}}
  g\left( \frac{2\mu+\omega}{vq} \right) -
 \theta_{\mu-\frac{\omega+vq}{2}}
 g\left(\frac{2\mu-\omega}{vq}\right)
 \right],
 \label{eq:imPI}
\end{align}
where $\theta_x$ denotes the step function satisfying 
$\theta_{x\ge 0}=1$ and $\theta_{x<0}=0$,
and $g(x)\equiv \log(x+\sqrt{x^2-1})$.
Figure~\ref{fig:imPI}(a) 
shows a 3-dimensional (3d) plot of $-{\rm Im} \Pi_\mu(q,\omega)$
as a function of $vq$ and $\mu$ 
when $\omega=0.2$ (eV), which
corresponds to the Debye frequency of carbon ($\omega_D$).
As indicated in Fig.~\ref{fig:imPI}(a),
a line node appears for $vq=\omega_D$.
In Eq.~(\ref{eq:imPI}), 
this line node is a critical line separating the
two terms, which are proportional to 
$\theta_{\omega-vq}$ and $\theta_{vq-\omega}$, and it can be shown that 
the first (second) term originates from the contributions of interband
(intraband) electron-hole pair creation.
For example, the $q=0$ phonon satisfies $\theta_{vq-\omega}=0$, 
and only the interband electron-hole pairs cause
spectral broadening.
The first term in Eq.~(\ref{eq:imPI}) leads to 
$-{\rm Im}\Pi_{\mu < \omega/2}(0,\omega)=\pi^2 \omega$
and ${\rm Im}\Pi_{\mu > \omega/2}(0,\omega)=0$,
which are consistent with
the behavior of the G band.~\cite{yan07,pisana07,das08nature,lazzeri06prl,ando06-ka} 
Contrastingly, for the phonon satisfying $vq>\omega$,
we can confirm that from 
the second term of Eq.~(\ref{eq:imPI})
spectral broadening 
is possible only when there is sufficient doping, namely
when $\mu>\frac{vq-\omega}{2}$ is satisfied.
A sharp step appears 
at $vq=2\mu +\omega_D$, as indicated in Fig.~\ref{fig:imPI}(a). 
In Fig.~\ref{fig:imPI}(b), 
we plot $-{\rm Im}\Pi_\mu(q,\omega_D)$ as a function of $\mu$
to show more clearly the $q$ dependence of the broadening.
It is seen that for $vq > 0.2$ [red curves],
$-{\rm Im}\Pi_\mu(q,\omega_D)$ is suppressed
when the Fermi energy is close to the Dirac point ($\mu<\frac{vq-0.2}{2}$),
and broadening appears when $\mu>\frac{vq-0.2}{2}$.

Because the Raman 2D band consists of two 
intervalley phonons satisfying $vq> \omega$,~\cite{saito02prl}
the suppressed broadening when $\mu\simeq 0$ also holds for the 2D
band. Das {\it et al.}~\cite{das08nature}
have shown that the 2D bandwidth sharpens when $\mu \simeq 0$.
A suppressed broadening 
of the 2D band ($2\omega_q \simeq 0.32$ eV)
has also been observed when $\mu \le 0.4$ eV
in a recent experiment reported by Chen {\it et al}.,~\cite{chen11}
from which we estimate the $q$ value to be $vq\simeq 0.96$ eV
using $0.4\simeq \frac{vq-0.16}{2}$.

The validity of this estimation ($vq\simeq 0.96$ eV)
can be further investigated by changing $q$.
In Fig.~\ref{fig:imPI}(c),
we show the plot of $-{\rm Im}\Pi_\mu(q,\omega_D)$
as a function of $q$ for different $\mu$ values.
For $vq\simeq 0.96$, increasing $q$ would cause
the broadening to decrease (increase) 
when $\mu=0.4$ (0.8) eV.
Because $vq$ is related to light excitation energy $E_L$
through momentum conservation,~\cite{saito02prl}
the broadening can also depend on $E_L$.
If we assume $E_L = \alpha vq$, $\alpha \simeq 1.6$
is obtained as a fitting parameter
because $E_L=1.58$ eV is used in the experiment.~\cite{chen11}
A similar $\alpha$ parameter value ($\alpha\simeq 1.3$) 
can be obtained by calculation.~\footnote{See Supplemental Material at [URL] for details.}

\begin{figure}[htbp]
 \begin{center}
  \includegraphics[scale=0.6]{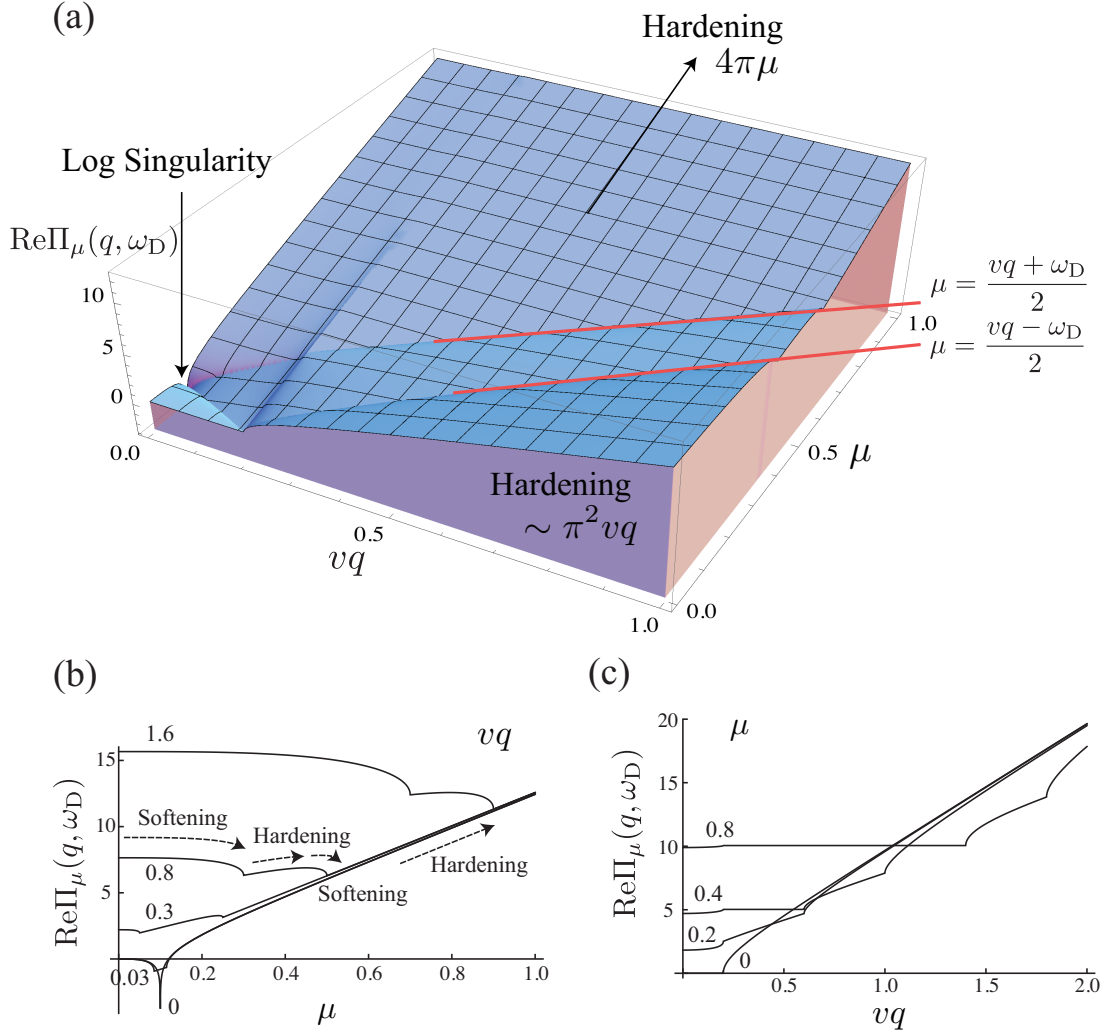}
 \end{center}
 \caption{
 (a) A 3d plot of ${\rm Re}\Pi_\mu(q,\omega_D)$.
 The variables $vq$ and $\mu$ are given in eV. 
 (b) The cross section of (a) for different $vq$ values,
 and (c) for different $\mu$ values.
 Note that ${\rm Re}\Pi_\mu(q,\omega)$ does not include the $q$
 dependence of the bare frequency.
 }
 \label{fig:rePI}
\end{figure}

A 3d plot of the real part of the self-energy,
${\rm Re}\Pi_\mu(q,\omega_D)$, 
is shown in Fig.~\ref{fig:rePI}(a).
The plot is based on 
the analytical expression of ${\rm Re}\Pi_\mu(q,\omega) $
given by~\footnote{See Supplemental Material at [URL] for derivation.}
\begin{align}
 &{\rm Re}\Pi_\mu(q,\omega) = 4 \pi \mu  \nn \\
 & \ +\pi \sqrt{\omega^2-v^2q^2}\theta_{\omega-vq}
 \left[-g\left(\frac{\omega+2\mu}{vq}\right)
 +\theta_{\frac{\omega-vq}{2}-\mu} g\left(\frac{\omega-2\mu}{vq}\right)
 +\theta_{\mu-\frac{\omega+vq}{2}}g\left(\frac{2\mu -\omega}{vq}\right)
 \right] \nn \\
 & \  
 + \pi \sqrt{v^2q^2-\omega^2}\theta_{vq-\omega} \left\{
 \theta_{\frac{vq-\omega}{2}-\mu}
 \left[\frac{\pi}{2} - \sin^{-1}\left(\frac{\omega+2\mu}{vq}\right)\right] 
 + \theta_{\frac{vq+\omega}{2}-\mu}\left[ \frac{\pi}{2}-\sin^{-1}\left(\frac{2\mu-\omega}{vq}
 \right)\right] \right\}.
\label{eq:rePI}
\end{align}
For the $q=0$ phonon, Fig.~\ref{fig:rePI}(a) shows that
the softening is maximum at $\mu=0.1$ (eV).
Equation~\ref{eq:rePI} is simplified in the limit of $q\to 0$, 
as
\begin{align}
 {\rm Re}\Pi_\mu(0,\omega) \simeq  4\pi
 \mu + \pi \omega \log\left|\frac{\omega-2\mu}{\omega+2\mu}\right|,
\end{align}
and the large softening is caused by the logarithmic singularity
at $\mu = \omega_D/2$.
This feature is exactly the same as the Kohn anomaly~\cite{kohn59} 
of the G band.~\cite{lazzeri06prl,ando06-ka,sasaki10-physicaE} 
Figure~\ref{fig:rePI}(a) shows that 
the logarithmic singularity is removed gradually 
as we increase $q$ from zero.
[The logarithmic singularity is obscured by charge inhomogeneity~\footnote{See Supplemental Material at [URL] for details.}]
It also shows that 
when $\mu$ is sufficiently large
the real part increases linearly with 
$\mu$ as ${\rm Re}\Pi_\mu(q,\omega_D)\simeq 4\pi \mu$ 
for an arbitrary $q$ value.
Interestingly, 
${\rm Re}\Pi_\mu(q,\omega_D)$ increases as we increase $q$ (or $E_L$),
even for a fixed $\mu$ value.
This feature is more clearly seen in Fig.~\ref{fig:rePI}(c), and 
suggests that the self-energy contributes to 
the dispersive behavior of the 2D (or D) band:~\cite{vidano81,matthews99,piscanec04}
the 2D band frequency increases linearly with $E_L$ 
($\partial \omega_{\rm 2D}/\partial E_L\simeq 100$ cm$^{-1}$/eV).~\cite{gupta09,casiraghi09} 
If we use $g_{\rm ep}=5$ cm$^{-1}$, which is obtained from 
the broadening data published by Chen {\it et al}.,~\cite{chen11}
the self-energy can account for $\sim$60\% of the dispersion
because $2\times {\rm Re}\Pi_{\mu\simeq 0}(q,\omega_q)\simeq 2g_{\rm ep}\pi^2
vq=2g_{\rm ep}\pi^2  E_L/\alpha$
and $2g_{\rm ep}\pi^2/\alpha=61.6$ cm$^{-1}$.

In Fig.~\ref{fig:rePI}(a),
for a fixed $vq$ that is larger than $\omega_D$,
${\rm Re}\Pi_\mu(q,\omega_D)$ undergoes two discontinuities
at $\mu=\frac{vq-\omega_D}{2}$ and $\frac{vq+\omega_D}{2}$.
A modest softening appears as $\mu$ approaches
$\frac{vq-\omega_D}{2}$ from zero.
This is consistent with the observations by 
Das {\it et al.}~\cite{das08nature}, Chen {\it et al}.~\cite{chen11},
and Araujo {\it et al.}~\cite{araujo12}
showing that the 2D band frequency remains almost constant (disregarding
a small modulation of about 8 cm$^{-1}$) 
when the Fermi energy is near the Dirac point.
On the other hand, the 2D band frequency exhibits a slight hardening of
$\sim2$ cm$^{-1}$ in the observation reported by Yan {\it et al.}~\cite{yan07}
We consider that the data actually show that the 2D band frequency
does not depend on doping 
because the observed small amount of hardening is within the
spectral resolution (2 cm$^{-1}$).~\cite{yan07}

\begin{figure}[htbp]
 \begin{center}
  \includegraphics[scale=0.9]{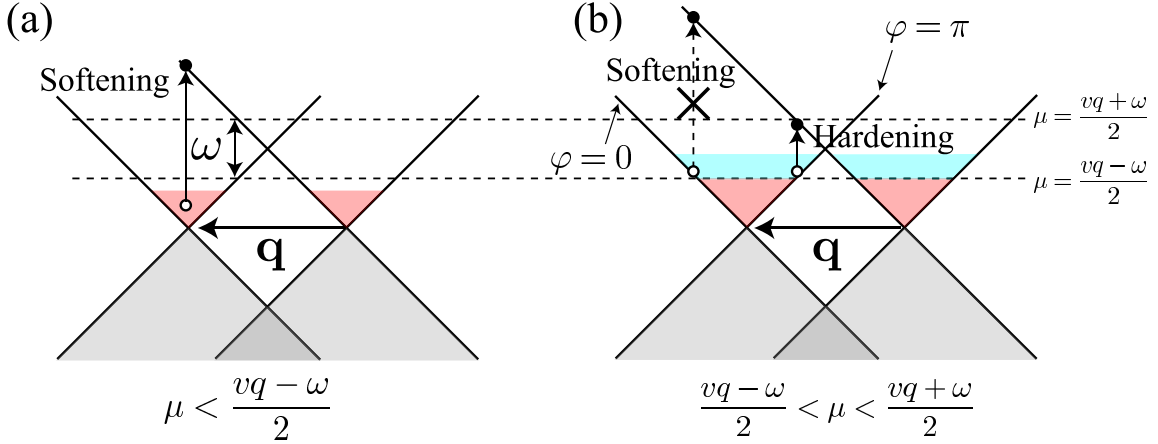}
 \end{center}
 \caption{
 The projection of the Dirac cones. 
 (a) Slight doping (red) causes a softening, 
 while (b) a heavy doping (blue) causes hardening as well as
 softening. 
 }
 \label{fig:PIdoping}
\end{figure}

As we increase $\mu$ further, 
${\rm Re}\Pi_\mu(q,\omega_D)$ undergoes slight hardening 
and subsequent softening 
until $\frac{vq+\omega_D}{2}$.
These features can also be seen in Fig.~\ref{fig:rePI}(b).
The discontinuities of ${\rm Re}\Pi_\mu(q,\omega)$ 
can be explained by the perturbation theory:
the energy correction by 
a virtual state with energy $\varepsilon$ is 
proportional to 
\begin{align}
 \frac{1}{\omega-\varepsilon}.
\end{align}
Because the sign of $(\omega-\varepsilon)^{-1}$ 
is positive (negative) when $\varepsilon<\omega$ ($\varepsilon>\omega$),
the lower (higher) energy electron-hole pair
makes a positive (negative) contribution to 
${\rm Re} \Pi_\mu(q,\omega)$.~\cite{sasaki10-physicaE}
Therefore, 
when $vq > \omega$,
softening is induced by the doped carriers 
since the energy of a virtual state
is approximately given by $vq$, which is larger
than $\omega$, and thus $1/(\omega-vq)<0$ is satisfied
[see Fig.~\ref{fig:PIdoping}(a)].
In fact, the energy $\epsilon$ corresponds to 
$v|{\bf k+q}|-vk$ in Eq.~(\ref{eq:Pi}) and $\epsilon\simeq vq$ 
when $k\simeq 0$.
The softening magnitude is tiny 
as shown in Fig.~\ref{fig:rePI}(a) and (b)
because the electron density vanishes at the Dirac point.
When $\mu=\frac{vq-\omega}{2}$, 
an intraband electron-hole pair with 
$\varepsilon \le \omega$ can start to be excited [see Fig.~\ref{fig:PIdoping}(b)], 
and this electron-hole pair causes hardening.
Note that some of the doped carriers satisfying 
$\frac{vq-\omega}{2} < \mu < \frac{vq+\omega}{2}$
contribute to the softening, and
the hardening is partly cancelled by the softening.
The details of the cancellation are determined by the $\varphi$
dependence of the electron-phonon coupling term, 
$1-ss' \frac{k+q\cos\varphi}{|{\bf k+q}|}$, 
in Eq.~(\ref{eq:Pi}).
Because the intraband transition satisfies $ss'=1$,
the matrix element vanishes when $\varphi=0$ and thus
hardening dominates softening
(unless $\omega$ is negligible compared with $vq$).
When $\mu$ approaches $\frac{vq+ \omega}{2}$,
the Pauli exclusion principle forbids 
the occurrence of some of the intraband transitions 
that contribute to the hardening, 
which accounts for the appearance of the softening.
For $\mu \ge \frac{vq+\omega}{2}$, the frequency exhibits hardening due
to the suppression of the softening
induced by interband ($ss'=-1$) virtual electron-hole pairs.

As we have seen, a phonon's self-energy 
can be very sensitive to the electron-phonon matrix element.
In particular, the sign of the coefficient of $ss'$ in 
\begin{align}
 1-ss' \frac{k+q\cos\varphi}{|{\bf k+q}|},
 \label{eq:Mintph}
\end{align} 
is critical in determining the behavior of the self-energy. 
In fact, if the minus sign is replaced with a plus sign as follows
\begin{align}
 1+ss' \frac{k+q\cos\varphi}{|{\bf k+q}|},
 \label{eq:pol}
\end{align} 
the corresponding self-energy 
exhibits a singularity at $vq=\omega$.
In addition,
the real part exhibits softening as $\sim -4\pi \mu$
when $vq>\omega$ 
because the matrix element is maximum when $\varphi=0$ and thus
softening dominates hardening. 
These features are in marked contrast to the fact that 
the imaginary part of the self-energy given by Eq.~(\ref{eq:Mintph})
exhibits a nodal structure at $vq=\omega$
and that the real part exhibits hardening as $\sim 4\pi \mu$.
The self-energy with Eq.~(\ref{eq:pol})
corresponds to the self-energy of the Coulomb potential
known as the polarization function,
and the singularity at $vq=\omega$
is important for plasmons in graphene.~\cite{wunsch06,hwang07}

\begin{figure}[htbp]
 \begin{center}
  \includegraphics[scale=0.7]{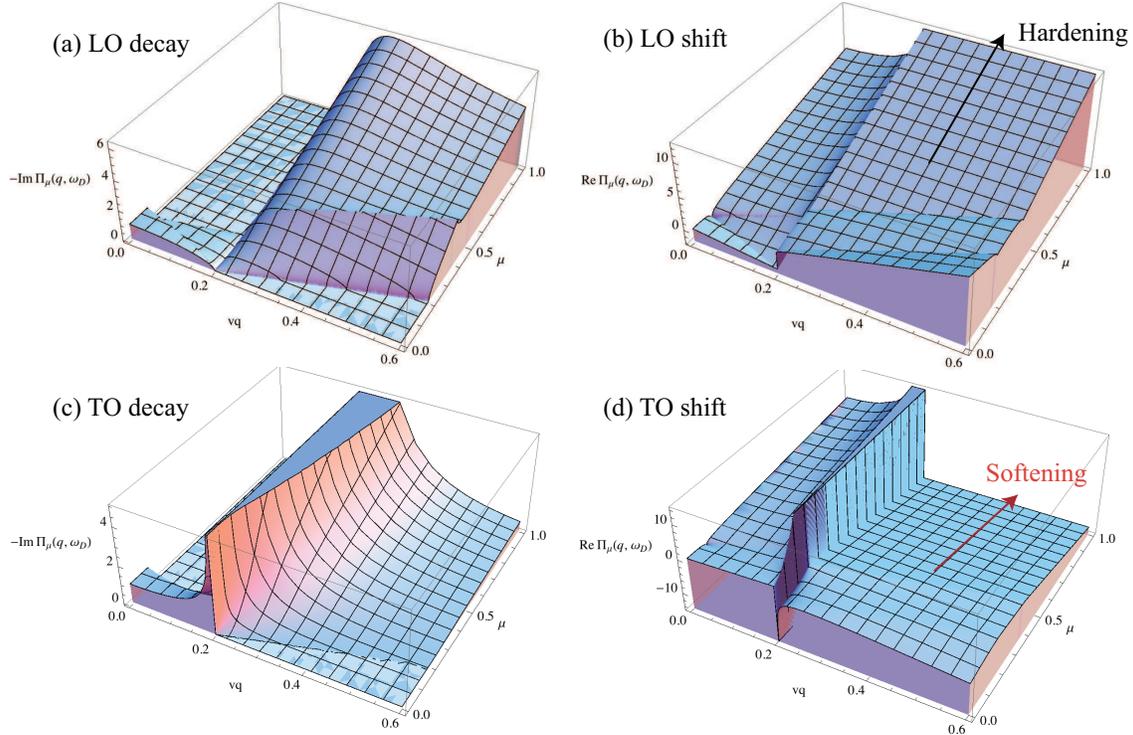}
 \end{center}
 \caption{
 3d plot of $\Pi_\mu(q,\omega_D)$ for an intravalley LO phonon (a,b)
 and TO phonon (c,d).
 (a,c) the imaginary part, and (b,d) the real part.
 The variables $vq$ and $\mu$ are given in eV. 
 }
 \label{fig:selfene_LO}
\end{figure}

Intravalley longitudinal optical (LO) and transverse optical (TO)
phonons are related to the intervalley phonon and plasmon.
The corresponding elements of the electron-phonon interactions are given
by~\footnote{See Supplemental Material at [URL] for derivation, ${\bf q}$ is measured from the $\Gamma$ point.}
\begin{align}
 &  1-ss' \frac{k+q\cos\varphi}{|{\bf k+q}|} + ss' \frac{2k\sin^2
 \varphi}{|{\bf k+q}|}, \ \ ({\rm LO}),
 \label{eq:iLO} \\
 &  1+ss' \frac{k+q\cos\varphi}{|{\bf k+q}|} - ss' \frac{2k\sin^2
 \varphi}{|{\bf k+q}|}, \ \ ({\rm TO}). 
 \label{eq:iTO}
\end{align}
The first two terms for the LO [TO] phonon are the same as
Eq.~(\ref{eq:Mintph}) [Eq.~(\ref{eq:pol})]. 
By constructing an analytical expression for the self-energy 
of the LO and TO phonons,~\footnote{See Supplemental Material at [URL] for derivation.}
we visualize the self-energies in Fig.~\ref{fig:selfene_LO}.
In Fig.~\ref{fig:selfene_LO}(a),
we show the imaginary part of the LO phonon.
A notable feature of Fig.~\ref{fig:selfene_LO}(a)
is that for $vq>\omega_D$, 
the broadening increases as $\mu$ is increased.
This is a sharp contrast to the broadening of the intervalley phonon,
which is suppressed for heavy doping (see Fig.~\ref{fig:imPI}(a)). 
For the real part shown in Fig.~\ref{fig:selfene_LO}(b),
a discontinuous feature caused by the last term in Eq.~(\ref{eq:iLO})
is clearly seen at $vq= \omega_D$.
Interestingly, 
the TO phonon has some similarities to the polarization
function: the existence of a singularity and
the frequency softening for $vq>\omega_D$, as shown in Fig.~\ref{fig:selfene_LO}(c) and (d). 
It is instructive 
to compare the self-energy of an intravalley LO phonon with that of a TO
phonon.
The self-energy of the LO phonon
is the same as that of the TO phonon for the $\Gamma$ point $q=0$,
and their difference is highlighted for nonzero $q$ values ($q >\omega/v$).
The difference between the LO and TO phonons
will be useful in allowing us to determine the optical phonon (LO or TO)
composing the D$'$ band.

Coulomb interactions among electrons which we did not consider in this
paper might make an effect on the self-energies of
phonons.~\cite{basko09-int}
Attaccalite et al. point out the importance of
vertex-corrections to electron-phonon coupling by electron-electron
interaction.~\cite{attaccalite10} 
However, their results concern the phonon at the exact K point,
which corresponds to the ${\bf q}=0$ phonon, 
and the ${\bf q}=0$ phonon has nothing
to do with the experimentally observed 2D band phonon. Our main results
remain unchanged even if we include such a correction, because our
approach - shifting the Dirac cones -  is based only on momentum
conservation and does not depend on any dynamical details. Moreover,
since we have determined the electron-phonon coupling using experimental
data, such effects, if any, are all included.

In conclusion,
employing a concept of shifted Dirac cones, 
we clarified that 
a phonon satisfying $vq > \omega$ does not decay into an
electron-hole pair when $\mu < (vq-\omega)/2$.
This is a general consequence 
that is independent of the details of electron-phonon coupling and 
that can be applied to both inter and intravalley phonons.
Based on the self-energy, which includes the effect of electron-phonon
coupling, we estimated the $q$ value of the 2D band at $vq\simeq 1$ (eV)
by referring to recent experimental data on the $\mu$ dependence of
broadening.
This value $vq\simeq 1$ (eV) also suggests that 
about 60\% of the dispersive behavior 
can be attributed to the self-energy.
Since $vq$ is proportional to $E_L$,
the $q$ dependence of the self-energy may be explored 
by using a tunable laser, without changing $\mu$ by controlling the gate
voltage.
For example, the Fermi energy position of graphene can be determined
from the $E_L$ dependence of the broadening.
Several anomalous features have been pointed out in
the self-energies for intravalley LO and TO phonons.
The differences between the LO and TO phonons 
will be useful for specifying the mode and $q$ value of the D$'$ band.

\bibliographystyle{apsrev}
%

\end{document}